\documentclass[twocolumn,prb,altaffilletter,amsmath,amssymb,amsfonts]{revtex4}
\usepackage[latin1]{inputenc}
\usepackage{graphicx}
\usepackage{amsmath}
\usepackage{latexsym}
\usepackage{epsfig}

\newcommand{\bR}{{\bf r}_1, ..., {\bf r}_N}

\begin{document}

\title{Pressure-energy correlations in liquids. V. Isomorphs in generalized Lennard-Jones systems.}

\author{Thomas B. Schr{\o}der}
\email{tbs@ruc.dk}
\author{Nicoletta Gnan}
\email{ngnan@ruc.dk}
\author{Ulf R. Pedersen}
\email{urp@ruc.dk}
\altaffiliation{Department of Chemistry, University of California, Berkeley, California 94720, USA}
\author{Nicholas P. Bailey}
\email{nbailey@ruc.dk}
\author{ Jeppe C. Dyre }
\email{dyre@ruc.dk}
\affiliation{DNRF Center ``Glass and Time'', IMFUFA, Dept. of Sciences, Roskilde University, P.O. Box 260, DK-4000 Roskilde, Denmark}

\date{\today}
\begin{abstract}
This series of papers is devoted to identifying and explaining the properties of strongly correlating liquids, i.e., liquids with more than 90\% correlation between their virial $W$ and potential energy $U$ fluctuations in the $NVT$ ensemble. Paper IV [N. Gnan {\it et al.}, J. Chem. Phys. {\bf 131}, 234504 (2009)] showed that strongly correlating liquids have ``isomorphs'', which are curves in the phase diagram along which structure, dynamics, and some thermodynamic properties are invariant in reduced units. In the present paper, using the fact that reduced-unit radial distribution functions are isomorph invariant, we derive an expression for the shapes of isomorphs in the $WU$ phase diagram of generalized Lennard-Jones systems of one or more types of particles. The isomorph shape depends only on the Lennard-Jones exponents; thus all isomorphs of standard Lennard-Jones systems (with exponents 12 and 6) can be scaled onto to a single curve. Two applications are given. One is testing the prediction that the solid-liquid coexistence curve follows an isomorph by comparing to recent simulations by Ahmed and Sadus [J. Chem. Phys. {\bf 131}, 174504 (2009)]. Excellent agreement is found on the liquid side of the coexistence, whereas the agreement is worse on the solid side. A second application is the derivation of an approximate equation of state for generalized Lennard-Jones systems by combining the isomorph theory with the Rosenfeld-Tarazona expression for the temperature dependence of potential energy on isochores. It is shown that the new equation of state agrees well with simulations.
\end{abstract}

\maketitle

\section{Introduction}

This is the fifth in a series of papers\cite{I,II,III,IV} investigating the properties of strongly correlating liquids,\cite{ped08a} i.e., liquids that have strong correlations between their constant-volume equilibrium fluctuations of potential energy, $U(t)$,  and  virial\cite{all87,han05} $W(t)\equiv -1/3 \sum_i {\bf r}_i \cdot {\bf \nabla}_{{\bf r}_i} U(\bR)$ where ${{\bf r}_i}$ is the position of particle $i$ at time $t$. As is well known, the average virial $W$ gives the configurational contribution to the pressure:
\begin{equation}\label{Wdef}
pV \,=\,
Nk_BT+W\,.
\end{equation}
Letting $\Delta$ denote instantaneous deviations from equilibrium mean values, the $WU$ correlation is quantified by the correlation coefficient $R$ (with $\left< ... \right>$ denoting equilibrium average):
\begin{equation}\label{R}
R \,=\,
\frac{\langle\Delta W\Delta U\rangle}
{\sqrt{\langle(\Delta W)^2\rangle\langle(\Delta U)^2\rangle}}\,.
\end{equation}
Perfect correlation gives $R=1$. As a pragmatic definition we have chosen ``strongly correlating liquids'' to designate liquids that have $R\ge 0.9$ in the $NVT$ ensemble (constant volume, $V$, and temperature, $T$). 

Strongly correlating liquids have simpler physics than liquids in general, an observation that has particular significance for the highly viscous phase.\cite{kau48,har76,bra85,gut95,edi96,ang00,alb01,deb01,bin05,sci05,dyr06} Thus it has been shown that strongly correlating viscous liquids to a good approximation have all eight frequency-dependent thermoviscoelastic response functions\cite{ell07,bai08c,chr08} given in terms of just one\cite{ped08b} (i.e., are single-parameter liquids in the sense of having dynamic Prigogine-Defay ratio\cite{ell07} close to unity\cite{ped08b,II,bai08c}). Strongly correlating viscous liquids moreover obey density scaling\cite{tol01,dre03,alb04,cas04,rol05} to a good approximation, i.e., their dimensionless relaxation time $\tilde\tau \equiv \tau \rho^{1/3}\sqrt{k_BT/m}$ (where $m$ is the average particle mass) depends on density $\rho=N/V$ and temperature as $\tilde\tau= F(\rho^\gamma/T)$.\cite{sch08b,cos08,cos09}

Paper I\cite{I} presented computer simulations of 13 different systems, showing that van der Waals type liquids are strongly correlating, whereas hydrogen-bonding liquids like methanol or water are not. Strongly correlating liquids include\cite{ped08a,ped08b,I,II,cos08}, for instance, the standard Lennard-Jones (LJ) liquid, the Kob-Andersen binary LJ (KABLJ) mixture, an asymmetric rigid-bond dumbbell model, a seven-site united-atom toluene model, and the Lewis-Wahnstr{\"o}m OTP model. 

Paper II\cite{II} analyzed the cause of $WU$ correlations with a focus on the LJ potential. The strong correlations were related to the well-known fact that an inverse power-law (IPL) pair potential, $v(r)\propto r^{-n}$ where $r$ is the distance between two particles, \cite{hoo72,hiw74,woo85,bar87,deb99,lan03,spe03,she03,ric05,bra06,cas06,hey07} implies perfect $WU$ correlation\cite{ped08a,II}, 

\begin{equation}
\Delta W(t)\, =\, \gamma \Delta U(t)\,
\end{equation}
with $\gamma=n/3$. Around the potential energy minimum, the LJ potential is well described by an ``extended'' inverse power-law potential (eIPL),\cite{II} $v_{\rm LJ}(r)\cong A r^{-n}+B+Cr$. 
At constant volume the linear term contributes little to the virial and potential-energy fluctuations: When one nearest-neighbor interatomic distance increases, others decrease in such a way that the sum is almost constant. Thus systems interacting via the LJ potential inherit strong $WU$ correlations from an underlying inverse power-law - they have a ``hidden scale invariance''.\cite{sch08b,III}

Paper III\cite{III} gave further numerical evidence for the explanation for strong $WU$ correlations presented in Paper II, and theoretical results were given on the statistical mechanics and thermodynamics of the hidden scale invariance that characterizes strongly correlating liquids. It was also shown that strong virial-potential energy correlations are present even in out-of-equilibrium situations - the hidden scale invariance is a property of the potential energy surface, not just of the equilibrium states. 

Paper IV\cite{IV} introduced the concept of ``isomorphs'' in the phase diagram of a strongly correlating liquid. Starting from a single assumption a number of  isomorph invariants were derived. In particular, structure and dynamics were shown to be invariant on isomorphs when reduced units are used.

In the present paper further simulation results supporting the isomorph predictions are presented using systems interacting with the multicomponent generalized LJ potential:  
\begin{equation}
  v_{ij}(r_{ij})=v_{ij}^{(m)}(r_{ij})+v_{ij}^{(n)}(r_{ij}),\,\,\label{Eq:GenLJ}
\end{equation}
where $v_{ij}^{(k)}(r_{ij})$ is an IPL potential acting between the two particles $i$ and $j$:
\begin{equation}
  v_{ij}^{(k)}(r_{ij})\equiv
  \varepsilon^{(k)}_{ij}\left({\sigma^{(k)}_{ij}}/{r_{ij}}\right)^k.
\end{equation}
For systems interacting via a generalized LJ potential, a prediction for the shape of the isomorphs in the $WU$ phase diagram is derived in section III and demonstrated to fit well to simulation results.\cite{simulations} Interestingly, the isomorph shape depends only on the exponents $m$ and $n$. Thus, e.g., all 12-6 LJ systems have the same isomorphs in the $WU$ phase diagram. Finally we briefly present two applications of the theory. One tests the Paper IV prediction that solid-liquid coexistence lines are isomorphs. The second application gives an approximate equation of state for systems interacting via generalized LJ potentials; this is arrived at by combining the present theory with Rosenfeld and Tarazona's expression for the isochoric temperature dependence of the potential energy.

\section{Isomorphs}

\subsection{Definition}

We term a microscopic configuration ``physically relevant'' if its influence on the thermodynamics and dynamics of the sustem is not {\it a priori} negligible (Paper IV). For instance, any configuration with very strong particle overlap is physically irrelevant; note, however, that even unlikely configurations like transition states are relevant. 

Two state points (1) and (2) with temperatures $T_1$ and $T_2$ and densities $\rho_1$ and $\rho_2$, respectively, are defined to be \emph{isomorphic} (Paper IV) if they obey the following: Any two physically relevant configurations of state points (1) and (2), $({\bf r}_1^{(1)}, ... , {\bf r}_N^{(1)})$ and $({\bf r}_1^{(2)}, ... , {\bf r}_N^{(2)})$, which trivially scale into one another, 
\begin{equation}\label{scaledef}
\rho_1^{1/3}\mathbf{r}_i^{(1)}
\,=\,\rho_2^{1/3}\mathbf{r}_i^{(2)}\,\,~~~(i=1,...,N)\,,
\end{equation}
have proportional configurational Boltzmann statistical weights:
\begin{equation}
e^{-U({\bf r}_1^{(1)}, ... , {\bf r}_N^{(1)})/k_BT_1}\, =\, C_{12}e^{-U({\bf r}_1^{(2)}, ... , {\bf r}_N^{(2)})/k_BT_2}\,. \label{Eq:IsomorphDEf}
\end{equation}
Here $U({\bf r}_1, ... , {\bf r}_N)$ is the potential energy function and it is understood that the constant $C_{12}$ depends only on the state points (1) and (2). 

The property of being isomorphic defines a mathematical equivalence relation on the set of state points. The corresponding equivalence classes are smooth curves in the phase diagram termed isomorphs or isomorphic curves.

\subsection{The approximate nature of isomorphs}

Equation (\ref{Eq:IsomorphDEf}) implies identity of the normalized Boltzmann probabilities for scaled relevant configurations of isomorphic state points. As detailed in Paper IV this identity implies that several quantities are invariant along an isomorph. Examples include the configurational entropy, the isochoric specific heat, N-particle distribution functions in reduced units (in particular, the radial distribution function(s)), reduced-unit dynamics (both for Newtonian and stochastic dynamics), normalized reduced-time autocorrelation functions, reduced-unit transport coefficients, etc. It was further shown in Paper IV that a jump from an equilibrium state to an isomorphic state point brings the system instantaneously to equilibrium (see also Ref. \onlinecite{gnan10}).

IPL potentials are Euler homogeneous functions, i.e., obey $U(\lambda{\bf r}_1, ... ,\lambda {\bf r}_N )=\lambda^{-n}U(\bR)$. Such purely repulsive potentials -- which do not describe real intermolecular interactions -- are the only potentials that have 100\% correlation between virial and potential energy fluctuations; this reflects the well-known IPL virial-theorem identity $W=(n/3)U$. Likewise, only IPL liquids have exact isomorphs; they satisfy Eq. (\ref{Eq:IsomorphDEf}) with $C_{12}=1$ for all configurations of two state points obeying $\rho_1^\gamma/T_1=\rho_2^\gamma/T_2$ where $\gamma=n/3$. 

Appendix A of Paper IV showed that strongly correlating liquids generally have isomorphs to a good approximation. Consequently, to a good approximation these liquids inherit a number of the exact invariants that IPL liquids have along their curves given by $\rho^\gamma/T={\rm Const.}$, namely all the IPL invariants that do not rely on the IPL identity $C_{12}=1$. Examples include the above mentioned thermodynamic, static, and dynamic isomorph invariants. IPL invariants that are not inherited by strongly correlating liquids in general include, e.g., the Helmholtz free energy over temperature, the potential energy over temperature, and the reduced-unit compressibility.

The fact that strongly correlating liquids have isomorphs to a good approximation should be understood as follows. Any liquid has curves in its phase diagram of constant excess entropy, curves of constant isochoric specific heat, curves of constant reduced relaxation time, curves of constant reduced-unit transport coefficients, etc. For a strongly correlating liquid these curves are almost identical, and they define its (approximate) isomorphs. Points on these isomorphs have approximately identical structure and approximately identical dynamics as probed, e.g., by normalized reduced-time-autocorrelation functions.

For clarity of presentation below we shall not repeatedly mention that the existence of isomorphs is  not exact for generalized LJ systems. Accordingly, we shall speak of isomorphs as a unique, well-defined concept. In other words: all properties of isomorphs derived below in generalized LJ systems would be exact if these systems did have exact isomorphs, e.g., fulfilled Eq. (\ref{Eq:IsomorphDEf}); since this is not the case, the properties are in principle -- as well as practice -- approximate.

\subsection{Generating isomorphs in simulations}

The structure in reduced units ($\tilde r \equiv \rho^{1/3}r$) is predicted to be invariant along an isomorph (Paper IV). The excess entropy, $S_ {\rm ex} = S - S_ {\rm ideal}$, depends only on structure and was shown in Paper IV also to be invariant along an isomorph (here and henceforth ``excess'' refers to the quantity in question in excess to that of an ideal gas with same density and temperature). In the below simulations we generate an isomorph as a set of state points in the phase diagram with constant $S_ {\rm ex}$. To change density and temperature keeping $S_ {\rm ex}$ constant, the following identity is used (Paper IV):
\begin{eqnarray}
 \gamma &\equiv& 
     -\left( \frac{\partial \ln T}{\partial \ln V}\right)_{\rm S_{ex}}  = 
     \frac{V \left( \frac{\partial S_{\rm ex}}{\partial V}\right)_T}
           {T \left( \frac{\partial S_{\rm ex}}{\partial T}\right)_V} = 
        \frac{V\beta_{\rm v}^{\rm ex}}{C_{\rm v}^{\rm ex}}  \label{Eq:GammaSex1}\,,
\end{eqnarray}
where $\beta_{\rm v}^{\rm ex} \equiv \left( \frac{\partial p_{\rm ex}}{\partial T}\right)_V = \frac{1}{V}\left( \frac{\partial W}{\partial T}\right)_V $ and a Maxwell relation was applied. Utilizing the standard fluctuation formulae this leads (Paper IV) to:

\begin{eqnarray}
 \gamma &=& 
     \left( \frac{\partial \ln T}{\partial \ln \rho}\right)_{\rm S_{ex}}  = 
       \frac{\left\langle \Delta U \Delta W \right\rangle}{\left\langle (\Delta U)^2\right\rangle}.  \label{Eq:GammaSex}
\end{eqnarray}

The procedure applied to generate isomorphs numerically can be summarized as:
1) An equilibrium $NVT$ simulation was performed at a given state point;
2) $\gamma$ was calculated from the fluctuations using Eq.~(\ref{Eq:GammaSex});
3) The density $\rho$ was changed slightly (of order 1\%), and $\gamma$ was used to calculate the corresponding change in temperature in order to keep $\rm S_{ex}$ constant (Eq. (\ref{Eq:GammaSex}));
4) A simulation at the new state point was performed, and the procedure was repeated.

\begin{figure}
\begin{center} 
 \includegraphics[width=0.45\textwidth]{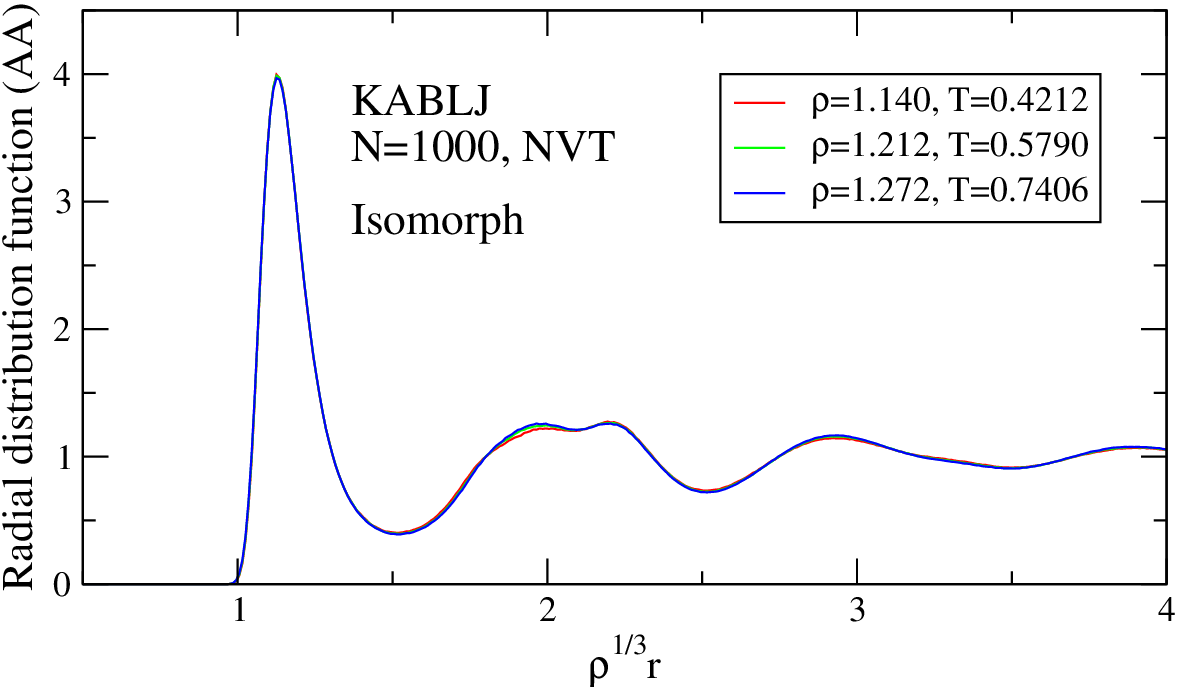}
 \includegraphics[width=0.45\textwidth]{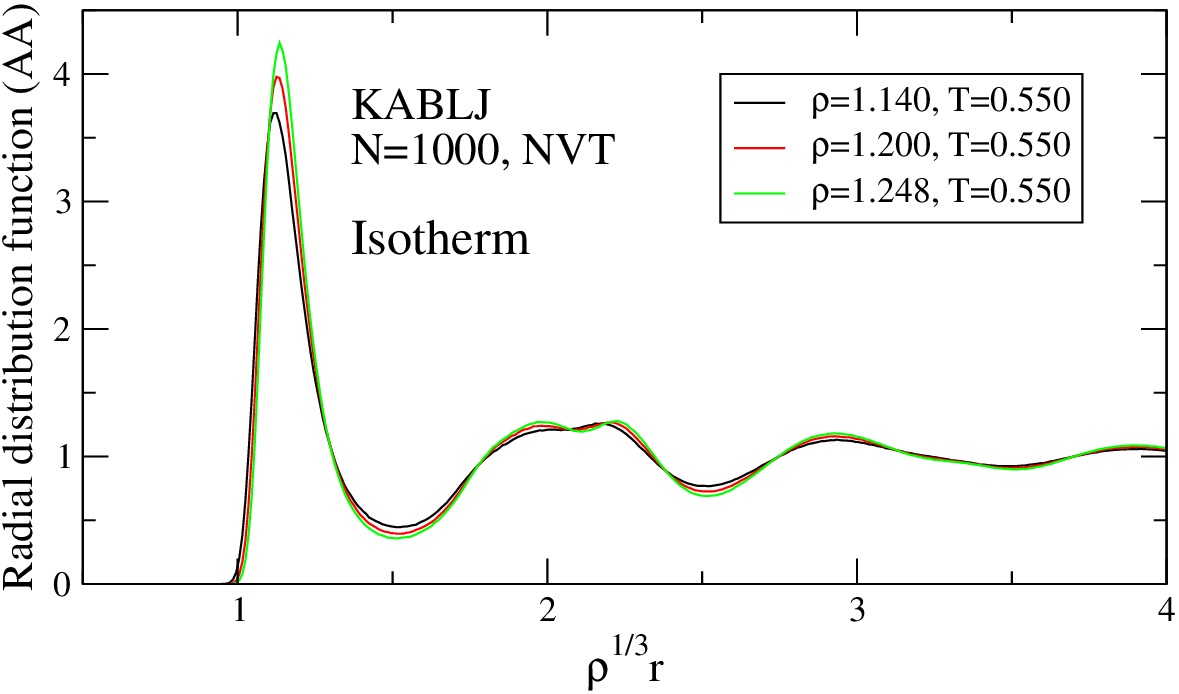}
\end{center}
\caption{a) Radial distribution function (AA particles) for three isomorphic state points of the Kob-Andersen binary LJ mixture \cite{KA} (KABLJ). b) Radial distribution function along an isotherm of the KABLJ mixture. A similar picture applies to the AB and BB radial distribution functions that are, however, slightly less isomorph invariant (Paper IV).}
\label{Fig:Structure}
\end{figure}

Figure~\ref{Fig:Structure}(a) demonstrates the isomorph invariance of the radial distribution function for the large (A) particles of the Kob-Andersen binary LJ mixture \cite{KA} (KABLJ).  For comparison,  Fig.~\ref{Fig:Structure}(b) shows the same on an isotherm with a similar (but smaller) density change. Clearly, structure is to a good approximation invariant on the isomorph, whereas this is not the case on the isotherm.   

\begin{figure}
\begin{center} 
 \includegraphics[width=0.45\textwidth]{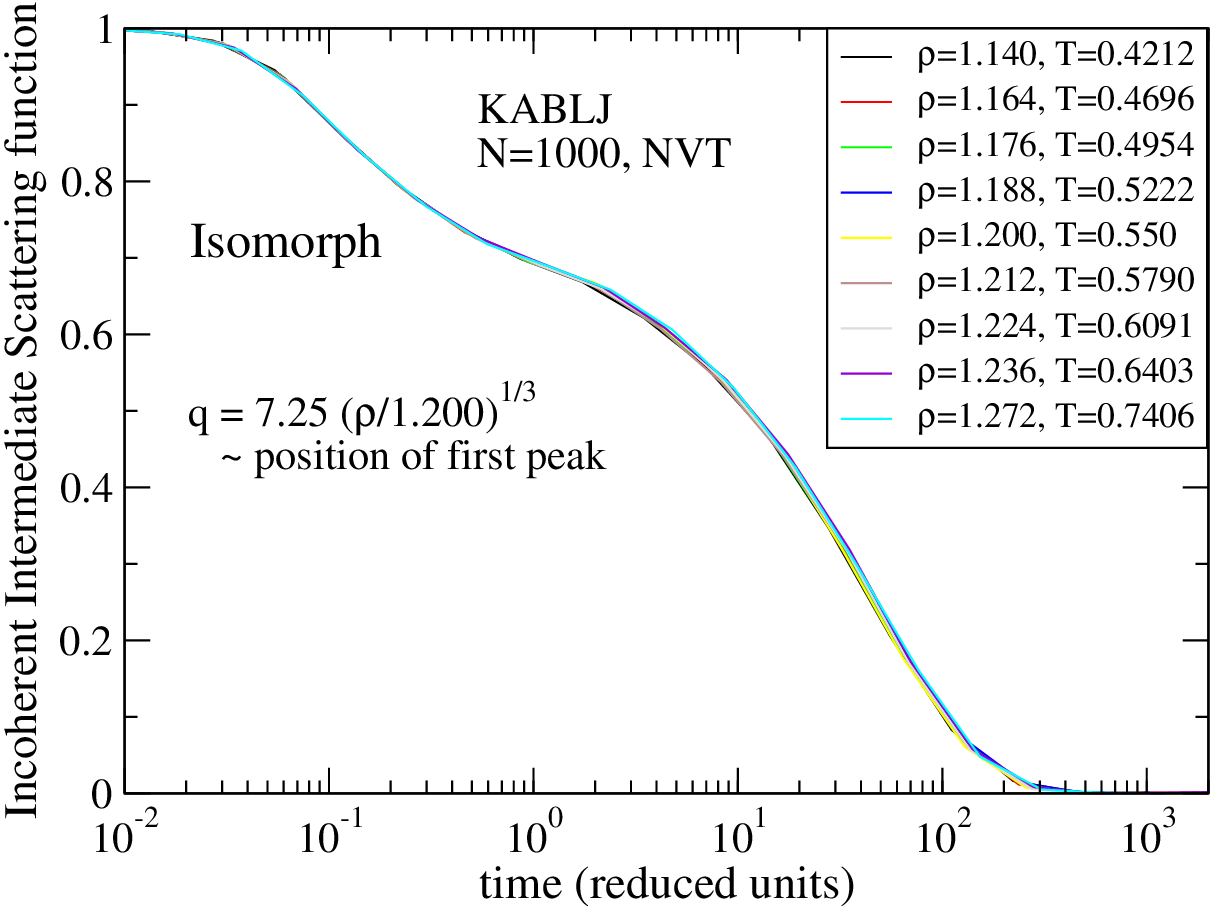}
 \includegraphics[width=0.45\textwidth]{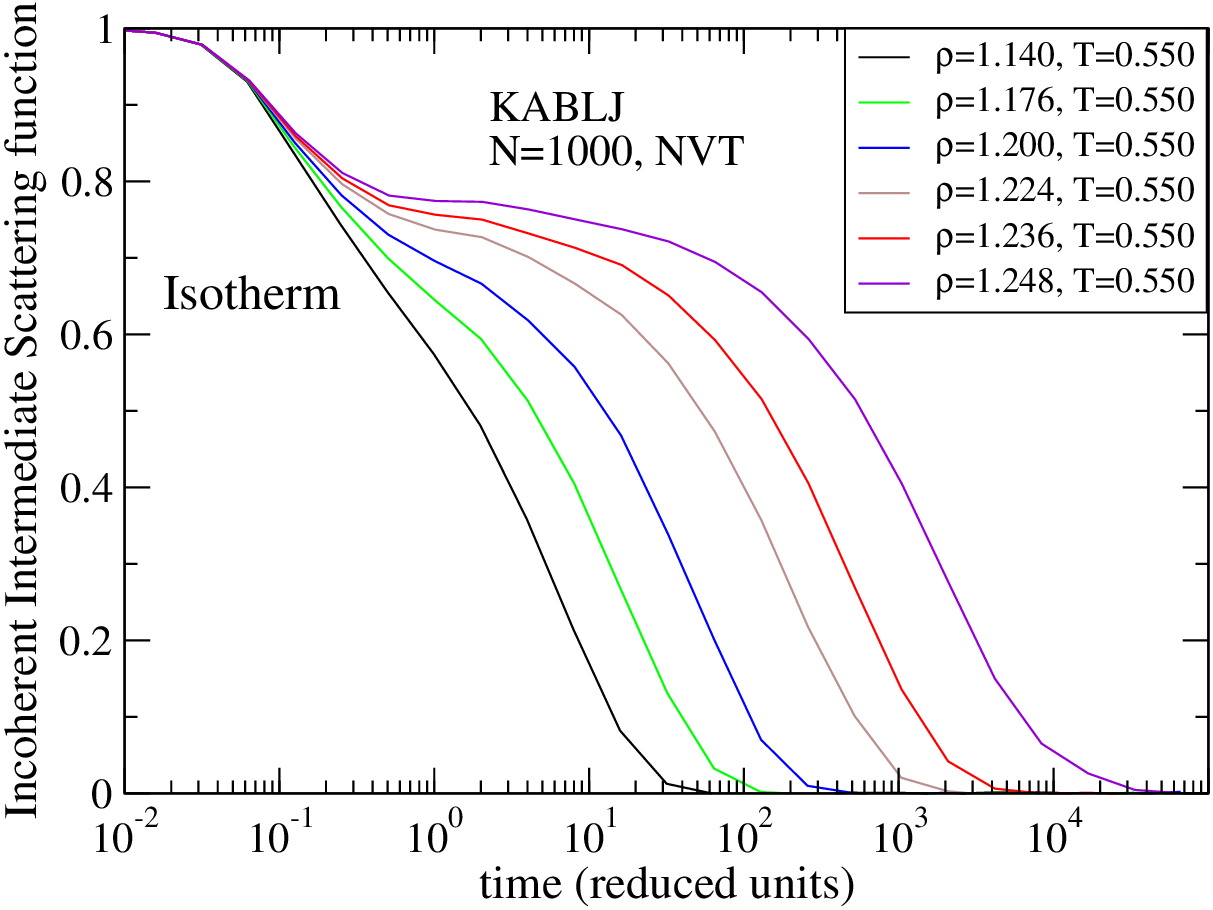}
\end{center}
\caption{a) Incoherent intermediate scattering function of the large (A) particles for isomorphic state points (density variation $11.6\%$) of the KABLJ mixture; time is given in reduced units: $\tilde t \equiv t \rho^{1/3}\sqrt{k_BT/m}$ and the q-vector is kept constant in reduced units. b) Same as in a), but along an isotherm (density variation $9.5\%$).}
\label{Fig:Dynamics}
\end{figure}

Taking the logarithm of Eq.~(\ref{Eq:IsomorphDEf}) shows that the potential energy surface in reduced units ($\tilde U \equiv U/k_BT$) is the same for two isomorphic state points, except for an additive constant. This constant does not influence the dynamics, and consequently the dynamics in reduced units is the same for the two isomorphic state points. Figure~\ref{Fig:Dynamics}(a) demonstrates the isomorph invariance of the incoherent intermediate scattering function of the A particles of the KABLJ mixture. For comparison,  Fig.~\ref{Fig:Dynamics}(b) shows the same property on the $T=0.55$ isotherm. Like the structure, the dynamics is also invariant to a good approximation  on the isomorph - both regarding  average relaxation time and the shape of the relaxation function; this is far from the case on the corresponding isotherm.

\begin{figure}
\begin{center} 
 \includegraphics[width=0.45\textwidth]{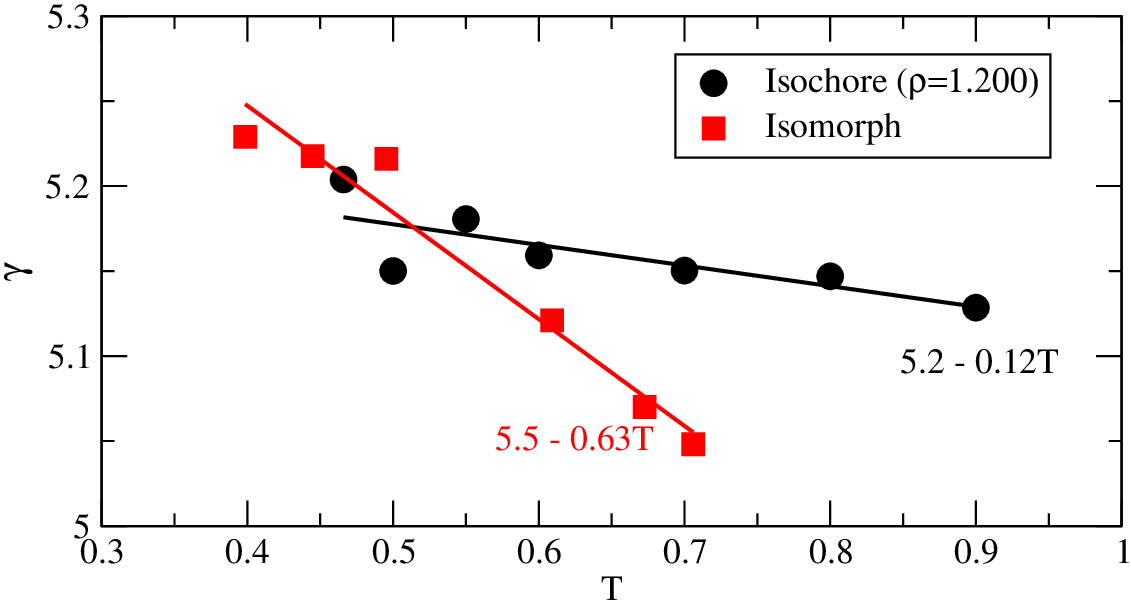}
\end{center}
\caption{$\gamma$ calculated from equilibrium fluctuations via Eq.~(\ref{Eq:GammaSex}) for the KABLJ system. Red squares: state points  belonging to the isomorph presented in Figs.~\ref{Fig:Structure} and \ref{Fig:Dynamics}. Black circles: isochoric state points. Straight lines are linear regression fits. Standard errors on slopes are 0.06 (isomorph) and 0.04 (isochore).
}
\label{Fig:GammaSex}
\end{figure}

The exponent $\gamma$ calculated from Eq.~(\ref{Eq:GammaSex}) generally depends on the state point. Figure~\ref{Fig:GammaSex} shows (as red squares) $\gamma$ along the isomorph of Figs. 1(a) and 2(a).Temperature changes by almost a factor of two, but $\gamma$ changes less than 5\%. Paper IV argued that for a system with good isomorphs $\gamma$ depends only on density. This is supported by Fig.~\ref{Fig:GammaSex} in which $\gamma$ on an isochore is shown to vary only $\sim 1$\% when temperature is changed by a factor of two (black circles).

\begin{figure}
\begin{center} 
 \includegraphics[width=0.45\textwidth]{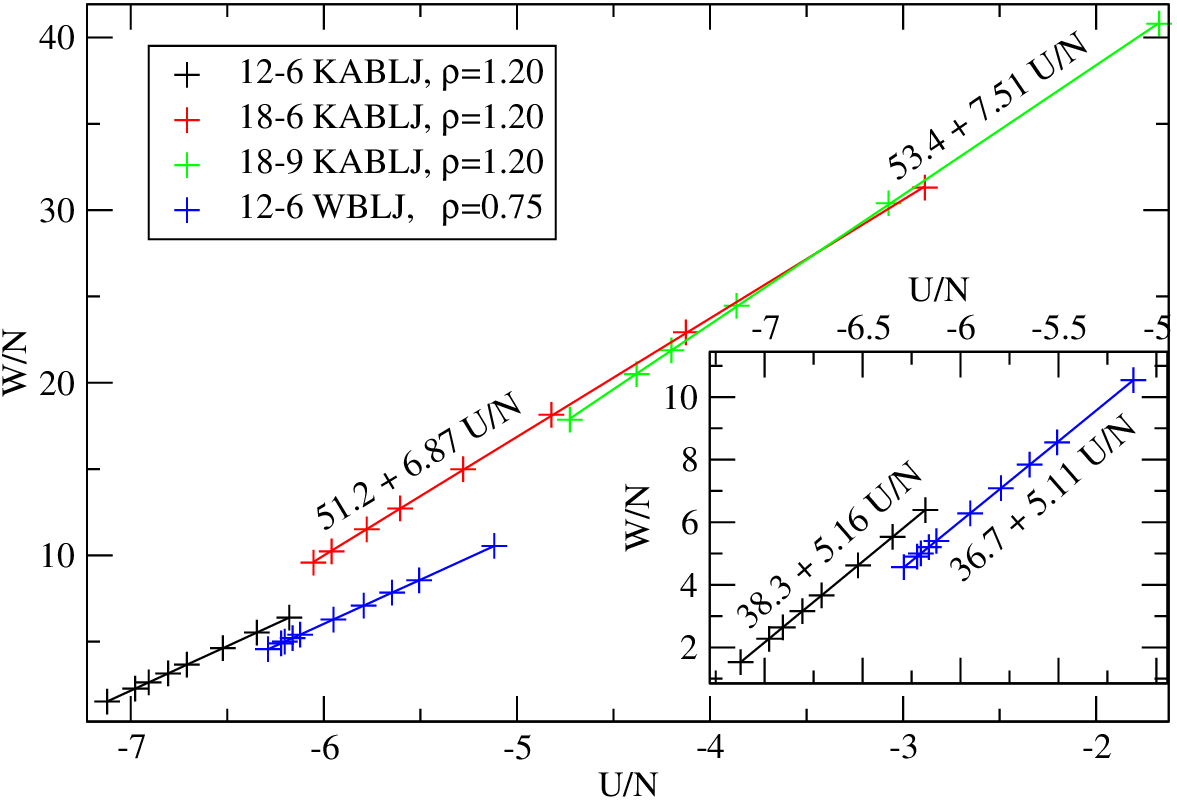}
\end{center}
\caption{Isochores for four different binary generalized LJ mixtures. 12-6 KABLJ is the Kob-Andersen binary 80:20 mixture\cite{KA} using the standard 12-6 LJ potential. 18-6 KABLJ is the same system with the generalized LJ potential ($m=18$, $n=6$, Eq.~(\ref{Eq:GenLJ})) and parameters chosen to place the minimum of the potential at the same distance and depth as in the 12-6 KABLJ system. Likewise, 18-9 KABLJ is using the exponents $m=18$ and $n=9$. Finally 12-6 WBLJ is the Wahnstr\"om 50:50 mixture\cite{wahn} using the standard 12-6 LJ potential. Straight lines are linear regression fits. The inset shows a zoom on the data for the two 12-6 LJ systems. 
For each system the simulated temperatures ranges from where non-exponential relaxation sets in ($\tau_\alpha \approx 10^0$) to a viscous state with two-step relaxation  ($\tau_\alpha \approx 10^3 - 10^5$).
}
\label{Fig:UWischores}
\end{figure}

It follows from Eq.~(\ref{Eq:GammaSex1}) that the slope of isochores in the $WU$ phase diagram is given  by $\gamma$: 
\begin{eqnarray}
 \gamma &=& 
      \left( \frac{\partial W}{\partial U}\right)_{\rm V}. \label{Eq:GammaSex3}
\end{eqnarray}
The insignificant change of $\gamma$ along isochores (black circles in Fig.~\ref{Fig:GammaSex}) means that these to a good approximation are straight lines in the $WU$ phase diagram. This is illustrated in Fig.~\ref{Fig:UWischores} for four different binary generalized LJ mixtures.

\section{Shape of isomorphs in the $WU$ phase diagram}

What is the shape of an isomorph in the $WU$ phase diagram? This section answers this question in generality for the multi-component generalized LJ potential. Recall that this potential (Eq.~(\ref{Eq:GenLJ})) is the sum of two IPL's. Correspondingly, the potential energy and virial can be expressed as sums of two IPL terms:
\begin{eqnarray}
 U &=& U_m + U_n, \qquad U_k \equiv \left< \sum_{i>j}v_{ij}^{(k)}(r_{ij})\right>. \label{eq:U}
\end{eqnarray}
For pair interactions the virial is given by\cite{all87} 
\begin{eqnarray}
 W \equiv -\frac{1}{3}\left< \sum_{i>j}r_{ij}v_{ij}^{'}(r_{ij})\right>, 
\end{eqnarray}
where the prime denotes the derivative with respect to $r_{ij}$. From this we get
\begin{eqnarray}
 W &=& \frac{m}{3}U_m + \frac{n}{3}U_n  \label{eq:W}.
\end{eqnarray}
For any point in the $WU$ phase diagram we can solve Eqs.~(\ref{eq:U}) and~(\ref{eq:W}) for $(U_m,U_n)$:
\begin{eqnarray}
 U_m &=& \frac{~~3W-nU}{m-n} \label{eq:Um}, \\
 U_n &=& \frac{-3W+mU}{m-n} \label{eq:Un}.
\end{eqnarray}

\begin{figure}
\begin{center} 
 \includegraphics[width=0.45\textwidth]{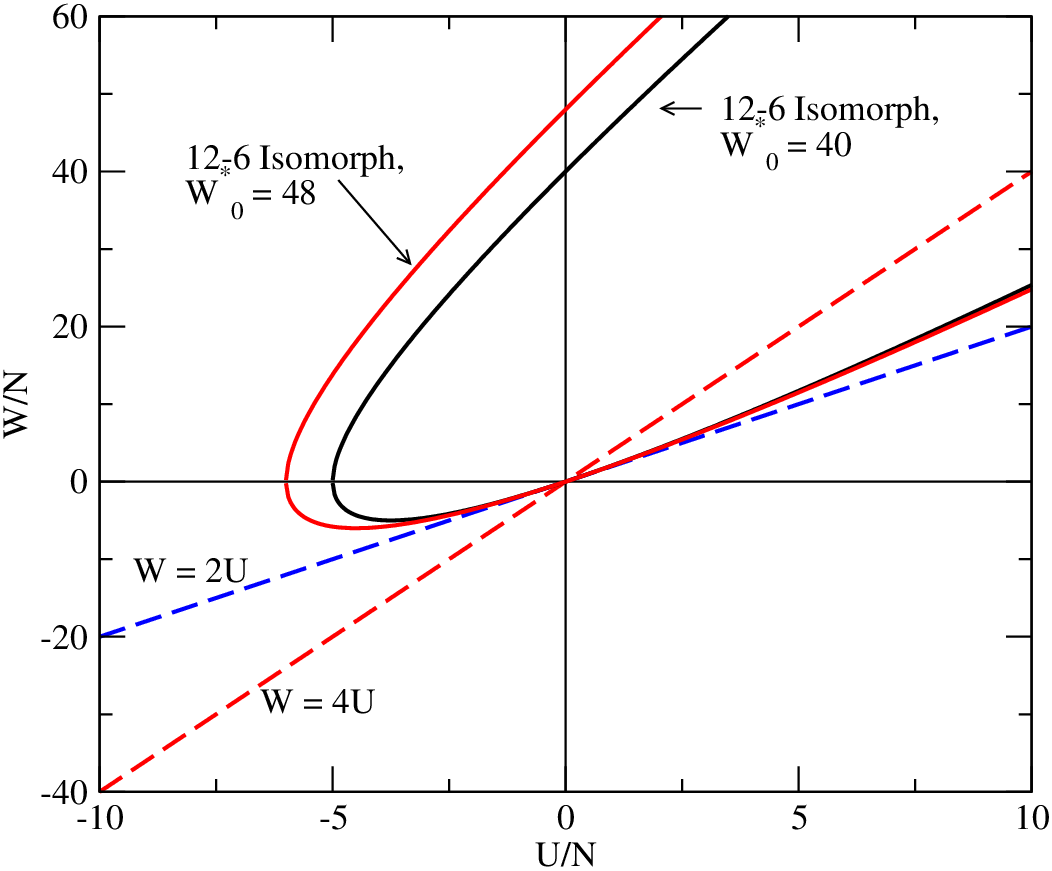}
\end{center}
\caption{
Schematic drawing of the $WU$ phase diagram for systems interacting via the standard 12-6 LJ potential. Included in this figure are also two theoretical isomorphs, plotted using Eq.~(\ref{eq:parabola}). The accessible points are those above both dashed lines (see text). The part of the isomorphs between the dashed lines in the $W,U>0$ quadrant are thus not accessible to the standard 12-6 LJ (they are  accessible to the purely repulsive 12-6 LJ potential with  $\epsilon_{ij}^{(6)}>0$.)} \label{Fig:GenericWU}
\end{figure}

These equations allow one to determine which regions of the $WU$ phase diagram are accessible to the system. Adopting the convention $m>n$, we restrict ourselves to the case $\epsilon_{ij}^{(m)}>0$ in order to have a repulsive core for all interactions. This implies  $U_m>0$ for all configurations, and therefore also for all thermodynamic averages. Combining this with Eq.~(\ref{eq:Um}) implies $W>\frac{n}{3}U$. If all interactions have an attractive part ($\epsilon_{ij}^{(n)}<0$) we have $U_n<0$, which in combinations with Eq.~(\ref{eq:Un}) implies $W>\frac{m}{3}U$. As an example, the region of the $WU$ phase diagram accessible to the standard 12-6 LJ potential is given by $W>4U$ and $W>2U$ (the later inequality being relevant when $U\leq 0$). See Fig.~\ref{Fig:GenericWU} for an illustration.

\subsection{Parametric description of isomorphs}

\begin{figure}
\begin{center} 
 \includegraphics[width=0.45\textwidth]{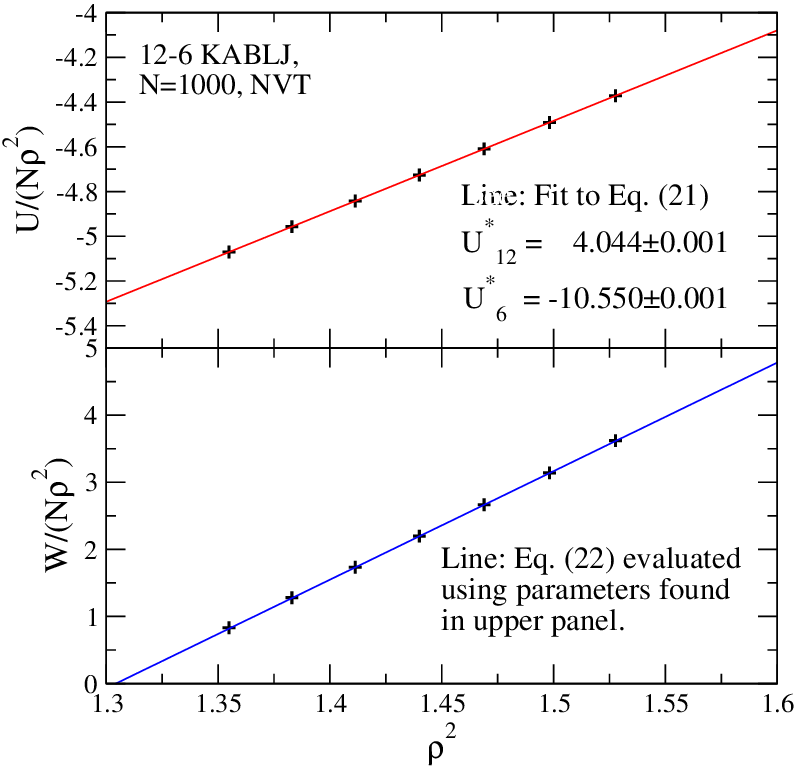}
\end{center}
\caption{
Test of 12-6 KABLJ isomorph (same as in Figs.~\ref{Fig:Structure} and \ref{Fig:Dynamics}). 
Top panel: U fitted to  Eq.~(\ref{Eq:IsomorphU_lin}) using $\rho_*=1$ (i.e., $\tilde \rho = \rho$). 
Bottom panel: W compared to Eq.~(\ref{Eq:IsomorphW_lin}), using the parameters  $U_m^*$ and $U_n^*$ found from $U$ (top panel).}
 \label{Fig:IsomorphUWtest}
\end{figure}

Along an isomorph the structure is invariant in reduced units (Fig.~\ref{Fig:Structure}). From this it follows that $U_k/\rho^{k/3}$ is invariant for any $k$ along an isomorph ($\tilde r \equiv \rho^{1/3}r$):
\begin{eqnarray}\label{dmath2}
 \frac{ U_k }{\rho^{k/3}}&
=&
\left< \sum_{i>j}\varepsilon^{(k)}_{ij}\left(\frac{\sigma^{(k)}_{ij}}{\tilde{r}_{ij}}\right)^{k}\right>  \\ &=&
 \sum_{i>j}\varepsilon^{(k)}_{ij}  \left({\sigma^{(k)}_{ij}}\right)^k                \left<{\tilde{r}_{ij}}^{-k}\right>\,.
\end{eqnarray}
If we let ``*'' denote a reference point, $(W_*, U_*)$, we can use Eqs. (\ref{eq:Um}) and (\ref{eq:Un}) to get $(U_m^*, U_n^*)$ and find for other state points on the same isomorph (where $\tilde{\rho}\equiv\rho/\rho_*$):

\begin{equation}\label{eq:scaled_U}
 U_k=\left(\frac{\rho}{\rho_*}\right)^{k/3}U_k^*=\tilde{\rho}^{k/3}U_k^*.
\end{equation}
Combining Eq.~(\ref{eq:scaled_U}) with Eqs.~(\ref{eq:U}) and (\ref{eq:W}) we obtain a parametric description of an isomorph in the $WU$ phase diagram:
\begin{eqnarray}
 U &=& \tilde{\rho}^{m/3}U_m^* + \tilde{\rho}^{n/3}U_n^*, \label{Eq:IsomorphU} \\
 W &=& \frac{m}{3}\tilde{\rho}^{m/3}U_m^* +\frac{n}{3}\tilde{\rho}^{n/3}U_n^*\label{Eq:IsomorphW} \,.
\end{eqnarray}
Once the numbers $U_m^*$ and $U_n^*$ have been determined from a reference state point using Eqs.~(\ref{eq:Um}) and (\ref{eq:Un}), the entire isomorph to which this state point belongs is thus traced out in the WU phase diagram using Eqs.~(\ref{Eq:IsomorphU}) and (\ref{Eq:IsomorphW}). Alternatively, given a collection of isomorphic state point generated, e.g., as described in section IIc, the predicted relation between $U$, $W$, and $\tilde\rho$ can be tested by linear regression:
\begin{eqnarray}
  \frac{U}{\tilde{\rho}^{n/3}} &=& \tilde{\rho}^{(m-n)/3}U_m^* + U_n^*, \label{Eq:IsomorphU_lin} \\
  \frac{W}{\tilde{\rho}^{n/3}} &=& \frac{m}{3}\tilde{\rho}^{(m-n)/3}U_m^* + \frac{n}{3}U_n^*\label{Eq:IsomorphW_lin} \,.
\end{eqnarray}
This is done in Fig.~\ref{Fig:IsomorphUWtest} for the 12-6 KABLJ isomorph presented in Figs.~\ref{Fig:Structure} and \ref{Fig:Dynamics}. The potential energies were fitted by linear regression by means of Eq.~(\ref{Eq:IsomorphU_lin}). Subsequently, the virials were compared to Eq.~(\ref{Eq:IsomorphW_lin}) using the parameters estimated from the potential energies (Eq.~(\ref{Eq:IsomorphU_lin})), i.e., without performing a new fit. 

\begin{figure}
\vspace{0.6cm}
\includegraphics[width=0.43\textwidth]{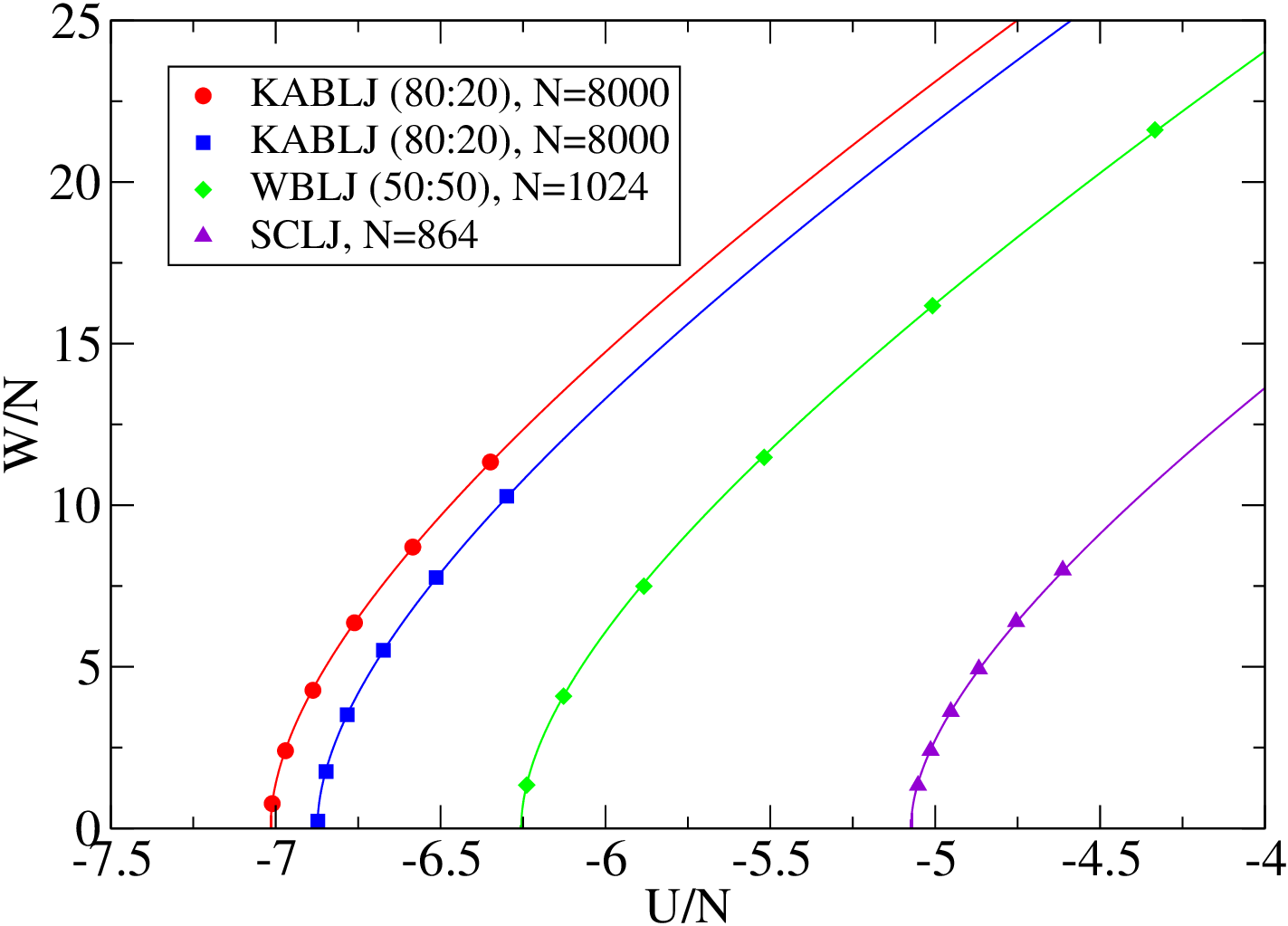}
\caption{
Four different 12-6 LJ isomorphs in the $WU$ phase diagram. The points give simulation results, the full curves the predictions of Eqs. (\ref{Eq:IsomorphU}) and  (\ref{Eq:IsomorphW}). Two isomorphs belong to the KABLJ system, the third result from simulations of Wahnstr{\"o}m binary LJ liquid\cite{wahn} (WBLJ), and the fourth (SCLJ) is the single component LJ liquid. 
}\label{Fig:Isomorph}
\end{figure} 

\begin{figure}
\includegraphics[width=0.45\textwidth]{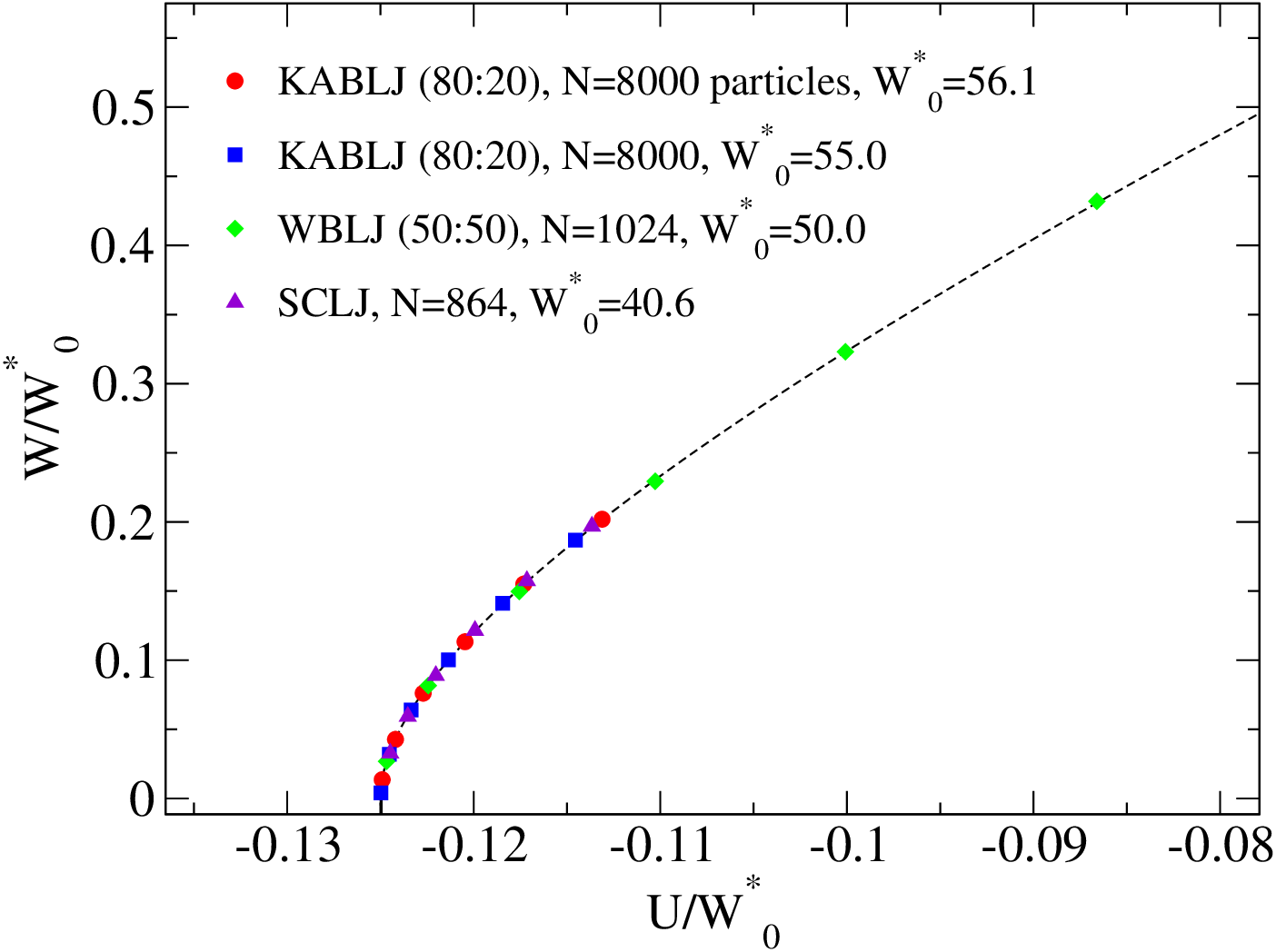}
\caption{
The ``master isomorph'': collapsing the four isomorphs from Fig.~\ref{Fig:Isomorph} by scaling with $W_0^*$ defined as the virial on each isomorph when $U=0$. The points give simulation results, the dashed curve the prediction of Eq. (\ref{eq:parabola}). }
\label{fig:figure10}
\end{figure}

\begin{figure}
\begin{center} 
 \includegraphics[width=0.45\textwidth]{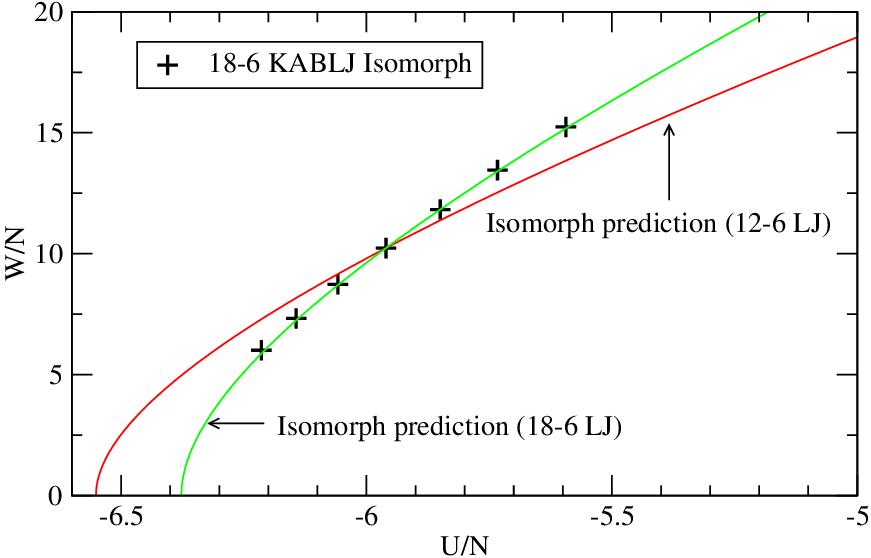}
\end{center}
\caption{
Isomorph of the 18-6 KABLJ liquid compared to the theoretical prediction (Eqs.~(\ref{Eq:IsomorphU}) and~(\ref{Eq:IsomorphW})). For reference, also the isomorph prediction for 12-6 LJ systems is included. }\label{Fig:Isomorph_18_6}
\end{figure}

We note here two important consequences of the above.
1) If the reference state point $(U_*,W_*)$ generates the isomorph $(U(\tilde\rho),W(\tilde\rho))$, then the reference state point $(aU_*,aW_*)$ generates the isomorph $(aU(\tilde\rho),aW(\tilde\rho))$. This means that all isomorphs for a given system have the same shape in the $UW$ diagram and can be scaled onto one another. 
2) The shape of the isomorphs depends only on the exponents, $m$ and $n$. This means that all 12-6 LJ systems have the same isomorphs in the $WU$ phase diagram -- they may be scaled onto a single ``master isomorph''. 

Figure~\ref{Fig:Isomorph} shows four isomorphs in the $WU$ phase diagram for three different 12-6 LJ systems.\cite{gromacs} Figure~\ref{fig:figure10} shows the same isomorphs scaled onto the 12-6 master isomorph, Figure~\ref{Fig:Isomorph_18_6} shows an isomorph for the 18-6 KABLJ system and compares it to a 12-6 LJ isomorph. Overall, the agreement between simulation results and predicted shapes is quite good.

\subsection{Eliminating the density parameter: The ``master isomorph'' equation}

Combining Eq.~(\ref{eq:scaled_U}) with Eqs.~(\ref{eq:Um}) and (\ref{eq:Un}) we get 
\begin{eqnarray}
 \tilde{\rho}^{m/3} = \frac{U_m}{U_m^*} = \frac{3W-nU}{3W_*-nU_*}, \\
 \tilde{\rho}^{n/3} = \frac{U_n}{U_n^*} = \frac{3W-mU}{3W_*-mU_*}. 
\end{eqnarray}
Eliminating density and rearranging, we find an invariant for the isomorph:
\begin{eqnarray}
 \frac{ \left( {W  -\frac{m}{3}U  }\right)^m }{ \left( {W  -\frac{n}{3}U  } \right)^n } =
 \frac{ \left( {W_*-\frac{m}{3}U_*}\right)^m }{ \left( {W_*-\frac{n}{3}U_*} \right)^n }. 
\end{eqnarray}
Choosing the reference point with zero potential energy $(U_*,W_*) = (0, W^*_{0})$ (this reference point is guaranteed to exist if all interactions have attraction), we get:
\begin{eqnarray}
 \frac{ \left( {W  -\frac{m}{3}U  }\right)^m }{ \left( {W  -\frac{n}{3}U  } \right)^n } =
  \left(W^*_{0}\right)^{m-n}.\label{eq:Invariant}
\end{eqnarray}
$W^*_{0}$ is a unique number identifying the isomorph. Equation~(\ref{eq:Invariant}) implies that the isomorph is the solution to 
\begin{eqnarray}
  \left( { \frac{W}{W^*_{0}}  -\frac{m}{3}  \frac{U}{W^*_{0}} }\right)^{m/n} =  \left( { \frac{W}{W^*_{0}}   -\frac{n}{3}  \frac{U}{W^*_{0}}  } \right)  
\end{eqnarray}
Here it can be seen directly that - as argued above - for a given set of exponents $(m,n)$ all isomorphs have the same shape, suggesting the name ``master isomorph''; $W^*_{0}$ is the parameter that determines the scale of each isomorph. For $m=2n$ (e.g., the standard 12-6 LJ potential) the solution to this (quadratic) equation is:
\begin{equation}\label{eq:parabola}
 2\frac{W}{W^*_{0}}={1 + 4\frac{n}{3}\frac{U}{W^*_{0}}\pm\sqrt{1 + 4\frac{n}{3}\frac{U}{W^*_{0}}}}{}.
\end{equation}
This has real solution(s) whenever  ${U}/{W^*_{0}}\geq -\frac{3}{4n}\,$, where equality corresponds to $W=0$. Figure \ref{Fig:GenericWU} plots Eq.~(\ref{eq:parabola}) with $n=6$ for two values of $W^*_{0}$.

\section{Applications}

\begin{figure}
\begin{center} 
 \includegraphics[width=0.235\textwidth]{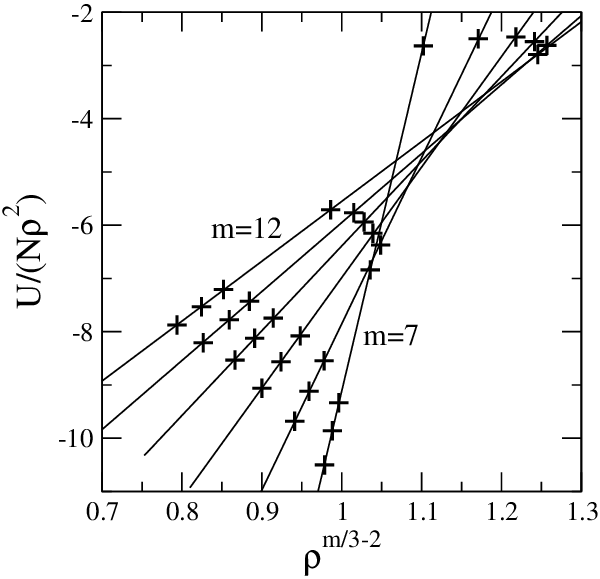}
 \includegraphics[width=0.235\textwidth]{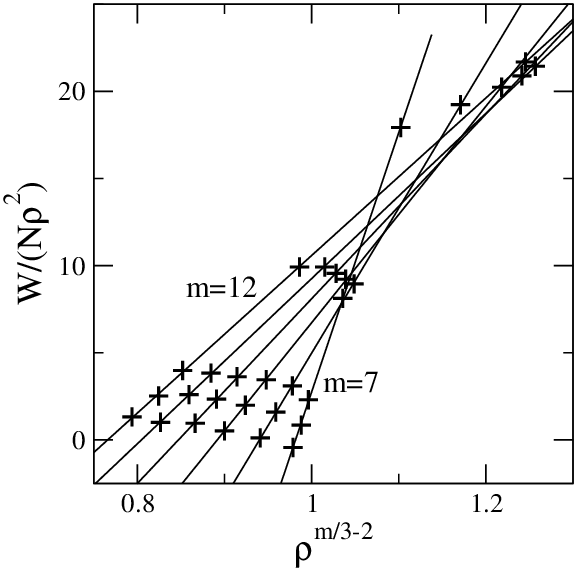}
 \includegraphics[width=0.3\textwidth]{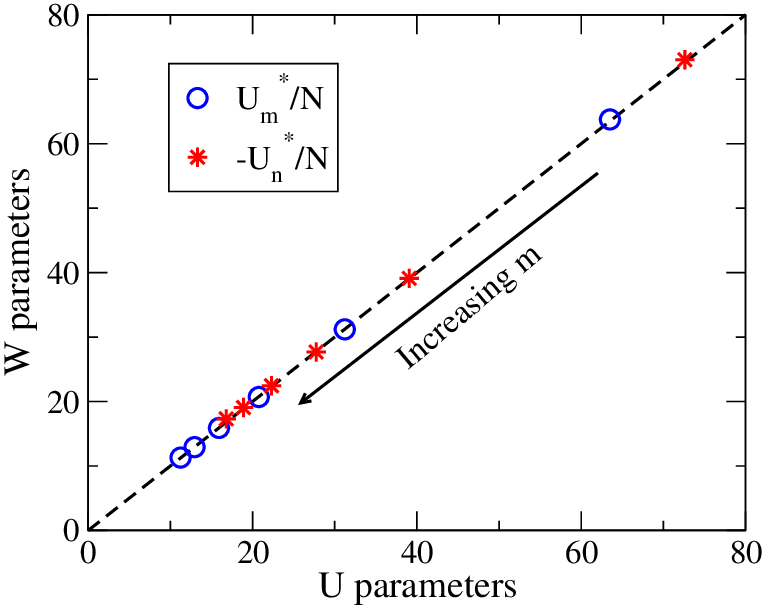}
\end{center}
\caption{Test of isomorph predictions for the liquid side of the solid-liquid coexistence curve of the single component generalized LJ potential ($n=6$). Left panel:  Eq.~(\ref{Eq:IsomorphU_lin}), right panel: Eq.~(\ref{Eq:IsomorphW_lin}); data points are simulation results from Ahmed and Sadus, \cite{Ahmed} straight lines are linear regression fits. The lower panel plots the fit-parameters from Eqs.~(\ref{Eq:IsomorphU_lin}) and (\ref{Eq:IsomorphW_lin}) against each other, demonstrating that the fits to energies and virials, respectively,  are consistent. 
}
\label{Fig:UWcoexist_lin}
\end{figure}

\begin{figure}
\begin{center} 
 \includegraphics[width=0.235\textwidth]{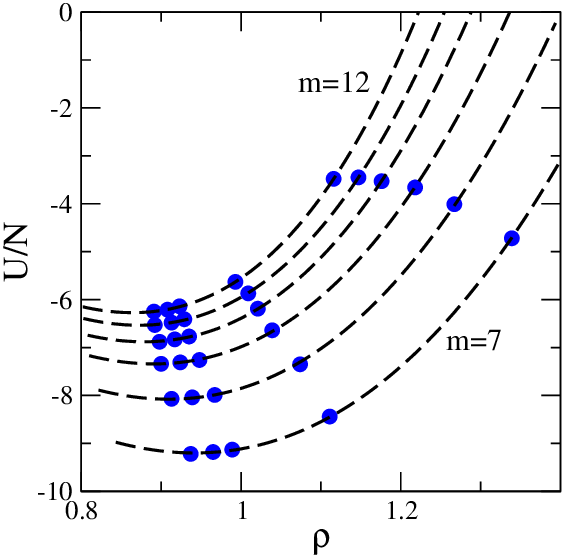}
 \includegraphics[width=0.235\textwidth]{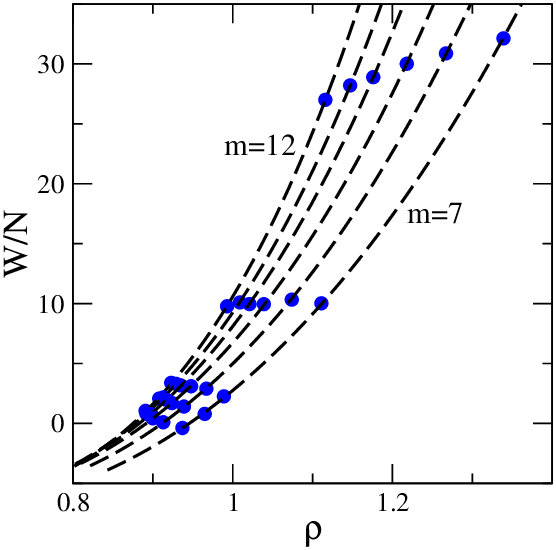}
\end{center}
\caption{Same data as in Fig. \ref{Fig:UWcoexist_lin}, plotting potential energy versus density (left panel) and virial versus density (right panel). Dashed lines: Isomorphic predictions (Eqs.~(\ref{Eq:IsomorphU}) and (\ref{Eq:IsomorphW})), using the parameters determined in   Fig.~\ref{Fig:UWcoexist_lin}.
}
\label{Fig:UWcoexist}
\end{figure}

\begin{figure}
\begin{center} 
 \includegraphics[width=0.45\textwidth]{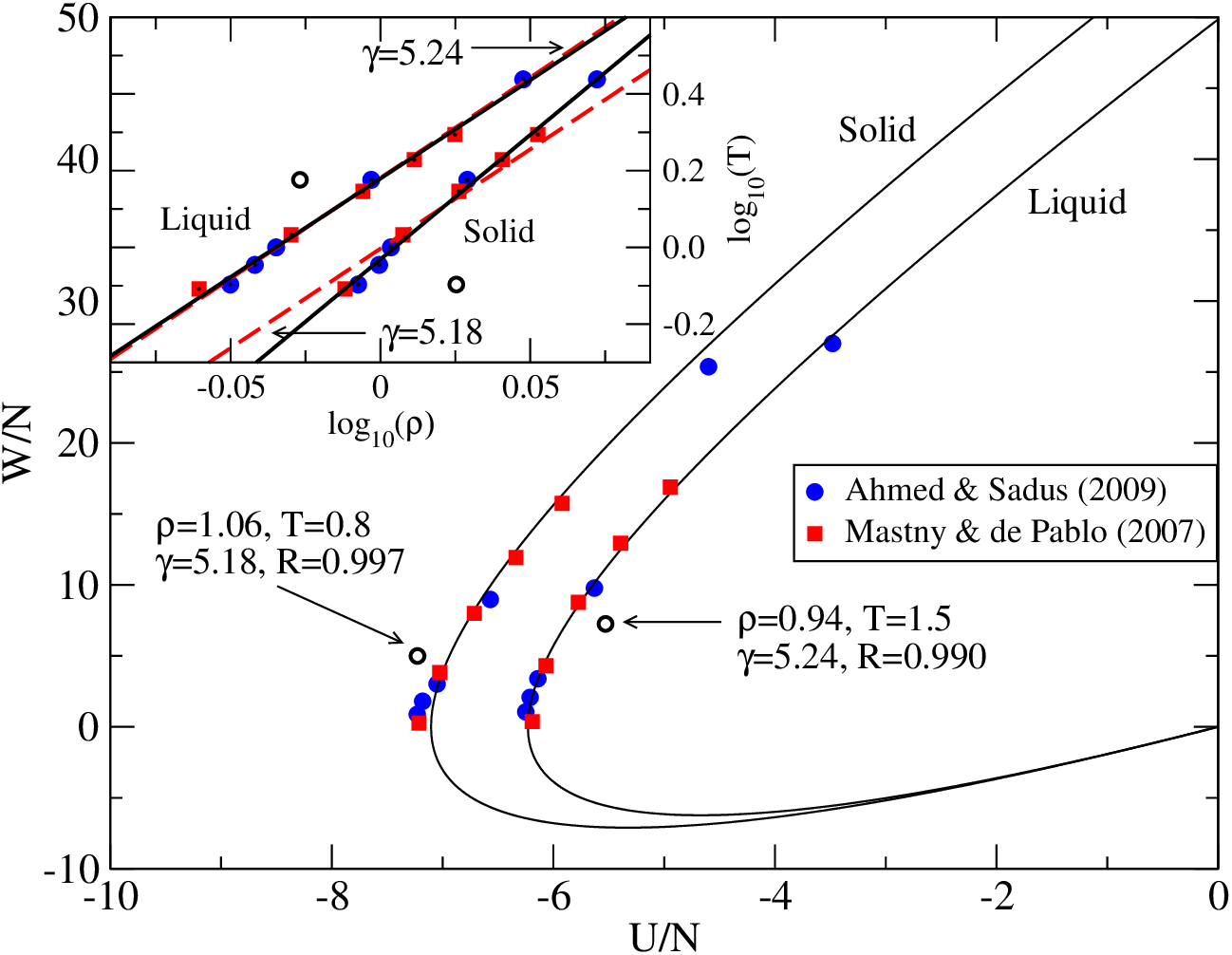}
\end{center}
\caption{Solid-liquid phase behavior for the single-component 12-6 LJ system. Main figure: Filled symbols mark the coexistence region in the WU diagram. Blue circles are from Ahmed \& Sadus \cite{Ahmed} and red squares are from Mastny \& de Pablo.\cite{Mastny} Full curves are the isomorph prediction (Eq.~(\ref{eq:parabola})) with $W_0^*/N=49.8$ and $56.8$ respectively. Open circles indicate results from simulations in the solid and liquid phase respectively. Both phases are (very) strongly correlating with $R\ge 0.99$. Inset shows the solid-liquid phase behavior in a $\rho,T$-diagram  with logarithmic axes. Dashed red lines are the isomorph prediction $\rho^\gamma/T=const$ using $\gamma$ found using Eq.~(\ref{Eq:GammaSex}) from the simulations of the two phases (open circles) giving  $\gamma=5.24$ and 5.18 respectively. Full black lines are free fits giving  $\gamma=5.13$ and 6.49 respectively.
}
\label{Fig:UWcoexistphase}
\end{figure}

\subsection{Solid-liquid coexistence}

In Paper IV it was argued briefly that Eq.~(\ref{Eq:IsomorphDEf}) predicts that solid-liquid coexistence lines are isomorphs. In more detail the argument goes as follows. First one notes that an isomorph cannot cross a solid- liquid coexistence line: Recall that the isomorph concept relates to equilibrium ensemble probabilities. For a pair of isomorphic state points all pairs of microscopic configurations related by  Eq.~(\ref{scaledef}) have identical Boltzmann probabilities according to Eq.~(\ref{Eq:IsomorphDEf}). Consequently, the two  isomorphic state points must also have the same macroscopic phase behavior. In other words, if an isomorph were to cross the coexistence line, part of the isomorph would have liquid microstates as the most likely and part of it would have mixed crystalline-liquid microstates as the most likely. This would violate the proportionality of Boltzmann statistical weights for scaled states, the property that defines an isomorph. 


Given that an isomorph cannot cross the solid-liquid coexistence line, the next point is to note that -- because all state points belong to some isomorph -- a curve infinitesimally close to the coexistence line must be an isomorph. This argument applies in the $W,U$ phase diagram, as well as in the density-temperature or the pressure-temperature phase diagrams. In the first two cases there is a multitude of isomorphs in the coexistence region; for instance one particular isomorph will be characterized by 53\% liquid and 47\% crystal, etc. In the pressure-temperature phase diagram these ``coexistence'' isomorphs collapse into one.

The prediction that the melting line, both slightly to the liquid side and slightly to the solid side, are isomorphs has a number of consequences and it sheds new light on well-known phenomenological melting rules (as well as exceptions to these rules for non-strongly correlating liquids, Paper IV). Thus, for instance the observation that the Lindemann melting criterion is pressure independent for a given liquid follows from the isomorph property of the melting line (whereas the existence of a universal Lindemann criterion does not follow). Indeed, as is easy to show from paper IV the isomorph theory implies Ross' ``generalized Lindemann melting law'' from 1969 \cite{Ross}, according to which the reduced configurational partition function of the crystalline phase
$Q^*=\int d\tilde{\bf R} \exp[-\beta(U(\tilde{\bf R})-U(0)] $ is invariant along the melting line (here $U$ is the potential energy, $U(0)$ is the potential energy of the crystal with all atoms located  exactly at their lattice sites, and the tilde denotes reduced coordinates. Further well-known melting rules, which the isomorph theory explains, include (Paper IV): The reduced viscosity, the reduced surface tension, the reduced diffusion constant, and the reduced heat conductivity are all invariant in the liquid phase along the melting line. Likewise, the reduced-unit static structure factor is invariant along the melting curve. These rules have been confirmed for several liquids in both experiment and simulation (see references of Paper V).


Like other isomorph predictions following from  Eq.~(\ref{Eq:IsomorphDEf}), the prediction that solid-liquid coexistence curves are isomorphs is only expected to be exact for systems that obey Eq.~(\ref{Eq:IsomorphDEf}) exactly, i.e., perfectly correlating systems. Indeed, it straightforward to show that in IPL systems the phase behavior depends only on $\rho^{n/3}/T$, and thus the co-existence line is an (exact) isomorph. The interesting question now is: how well do strongly correlating systems, e.g., generalized LJ systems, follow the prediction? Note that we do not expect the liquid-gas co-existence lines to be isomorphs (and indeed they are not); the gas-phase is not strongly correlating and thus  Eq.~(\ref{Eq:IsomorphDEf}) cannot be fulfilled in the relevant part of configuration space. In contrast, at the solid-liquid coexistence both phases involved are strongly correlating for LJ systems (Paper I), and we thus in the following test to what degree the isomorph prediction is fulfilled.

Ahmed and Sadus used in a recent paper \cite{Ahmed} a novel numerical method for determining solid-liquid coexistence. They reported results for the generalized LJ potential with the repulsive exponent $m$ varying from 12 to 7 and fixed attractive exponent $n=6$ (Ahmed and Sadus termed the repulsive exponent $n$, not $m$; we keep, however, the above notation). Figure \ref{Fig:UWcoexist_lin} shows the state points reported by Ahmed and Sadus for the liquid side of the solid-liquid coexistence curve compared to the isomorphic predictions as expressed in Eqs.~(\ref{Eq:IsomorphU_lin}) and (\ref{Eq:IsomorphW_lin}). The lower panel shows that the parameters found by fitting to  Eqs.~(\ref{Eq:IsomorphU_lin}) and (\ref{Eq:IsomorphW_lin}) respectively are consistent. Figure~\ref{Fig:UWcoexist} shows the same data plotting, respectively, potential energy (left panel) and virial (right panel) versus density.  Figure \ref{Fig:UWcoexistphase} shows the 12-6 LJ  solid-liqud phase behavior in the WU-diagram and in the $\rho T$-diagram (inset), including here also data from Mastny and de Pablo.\cite{Mastny}

The prediction that solid-liquid coexistence lines are isomorphs agrees well on the liquid side with the simulation results of Ahmed and Sadus\cite{Ahmed} as well as with the results for 12-6 LJ from Mastny and de Pablo\cite{Mastny}. On the solid side the agreement is less convincing (see Fig.~\ref{Fig:UWcoexistphase}); in particular the logarithmic derivative in the $\rho T$ diagram (slope in the inset) differs by 25\%. At present we do not have an explanation for this. It might be related to inter-particle distances longer than the first peak of the radial distribution function playing a bigger role for the phase behavior on the solid side\cite{Stillinger} than on the liquid side.

\begin{figure}
\begin{center} 
 \includegraphics[width=0.45\textwidth]{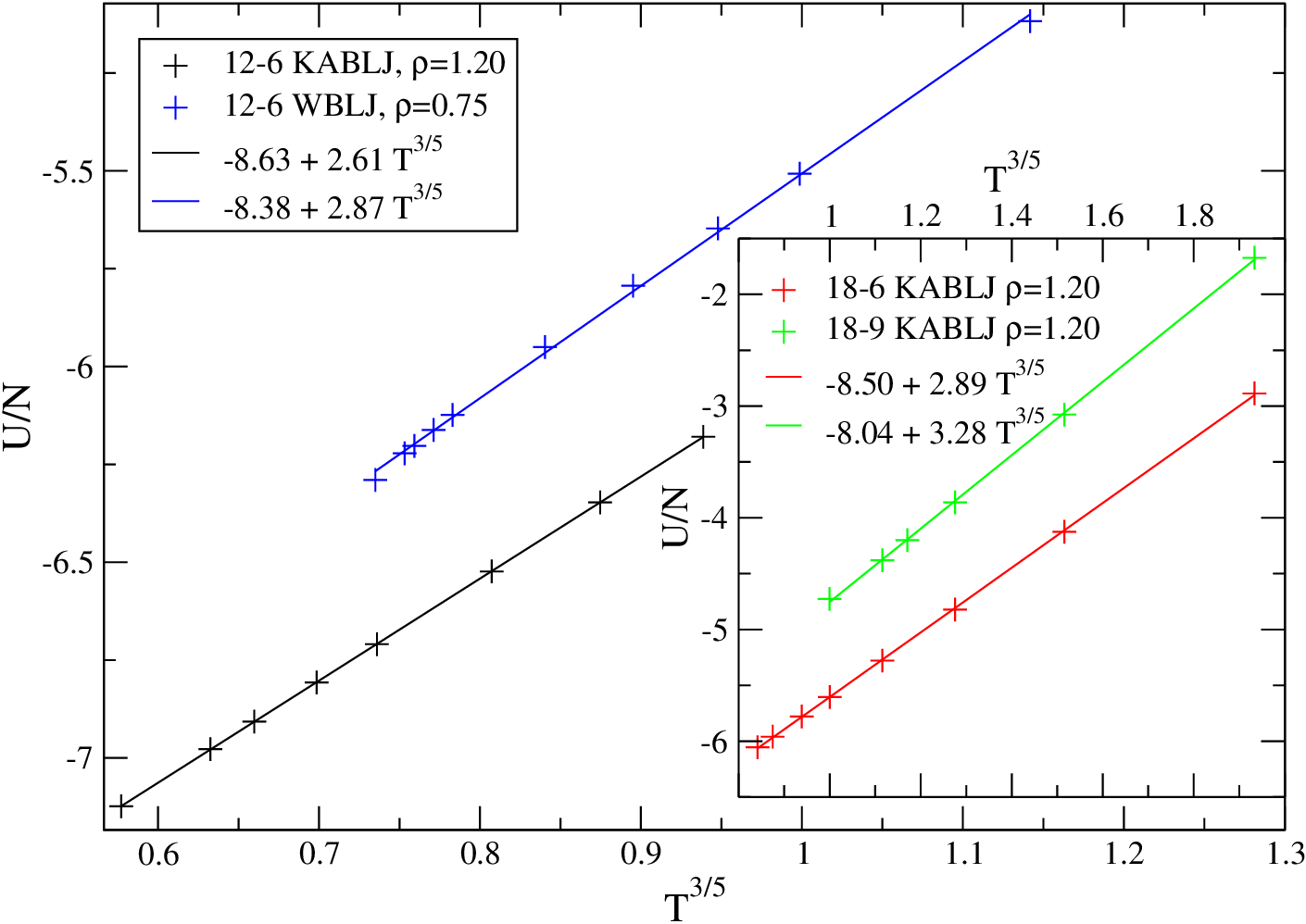}
\end{center}
\caption{Test of Rosenfeld-Tarazona scaling (Eq.~(\ref{Eq:URosenfeld})) plotting potential energy versus $T^{3/5}$ on isochores for the systems investigated in Fig. \ref{Fig:UWischores}.
}
\label{Fig:URosenfeld}
\end{figure}

\begin{figure}
\begin{center} 
 \includegraphics[width=0.45\textwidth]{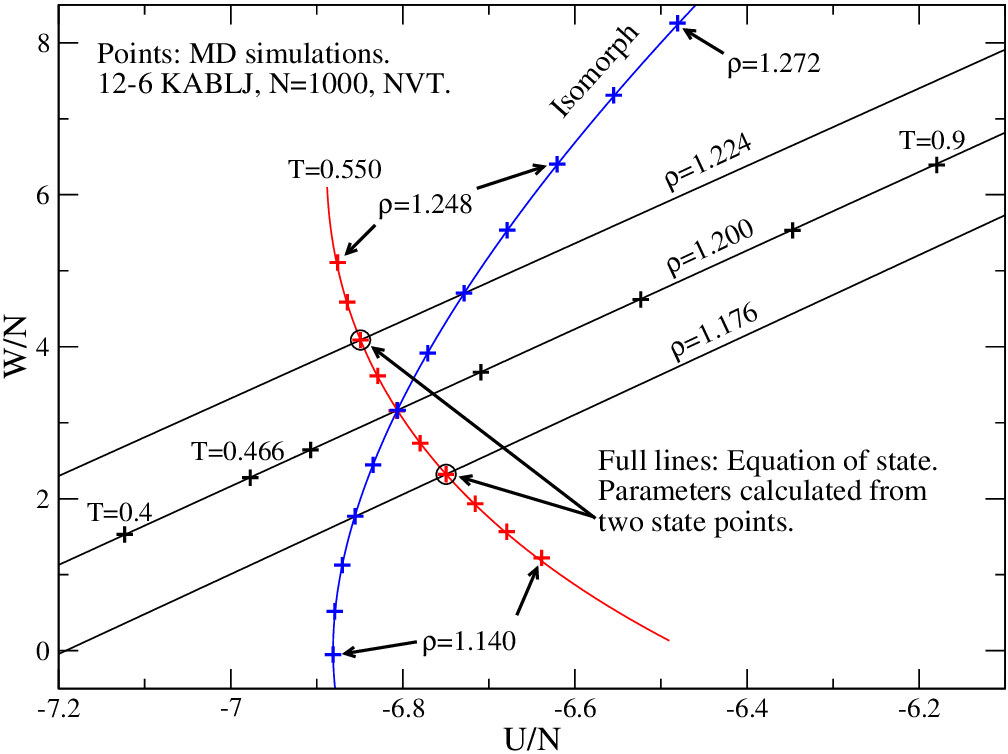}
 \includegraphics[width=0.45\textwidth]{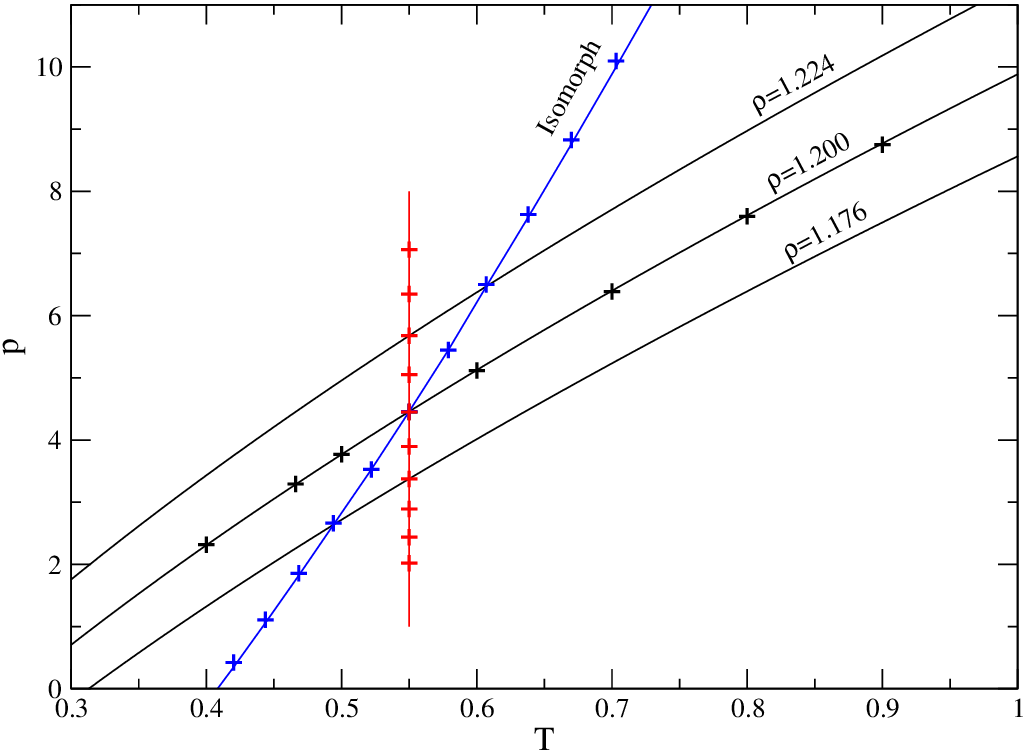}
\end{center}
\caption{Upper panel: Test of the equation of state in the $WU$ phase diagram using $\rho=1.200$ as reference density. Curves and lines: theoretical predictions (Eqs.~(\ref{Eq:Ustar}) to (\ref{Eq:isomorph_scaling_W})) for three isochores, an isomorph, and an isotherm. Data points: simulations of the 12-6 KABLJ mixture. The parameters of the equation of state were found from the two state points indicated by circles, giving $(W_0/N,\gamma,U_0/N,\alpha/N) = (38.4, 5.17, -8.63, 2.61)$. By comparison, the corresponding parameters found by fitting to the $\rho=1.200$ isochore in Figs.~\ref{Fig:UWischores} and \ref{Fig:URosenfeld} are $(38.3, 5.16, -8.63, 2.61)$, illustrating the robustness of the fitted equation of state. Applying the fitting procedure to the simulated state points on the  $T=0.550$ isotherm gives the parameters (38.5, 5.19, -8.63, 2.61).
Lower panel: Same data as above in a $pT$ diagram.
}
\label{Fig:UWstate}
\end{figure}

\subsection{An approximate equation of state for generalized LJ systems}

Equations~(\ref{Eq:IsomorphU}) and (\ref{Eq:IsomorphW}) give the potential energy and the virial as functions of density along an isomorph. The temperature variation along the isomorph can be found by integrating Eq.~(\ref{Eq:GammaSex}). Since $\gamma$ generally changes only little (Fig.~\ref{Fig:GammaSex}), in many situations it is a good approximation to assume it constant, leading to the relation $T=\tilde\rho^\gamma T_*$ along an isomorph, where $T_*$ is the temperature at the reference point. Taken together Eqs.~(\ref{Eq:IsomorphU}), (\ref{Eq:IsomorphW}), and (\ref{Eq:GammaSex}) thus imply that from a single reference point $(\rho_*,T_*,U_*,W_*)$ we have a prediction for $(\rho,T,U,W)$ on the isomorph to which the reference point belongs.

Figure \ref{Fig:UWischores} demonstrated that in the $WU$ phase diagram, isochores to a good approximation are straight lines with slope $\gamma$, i.e., that one can write $W(\rho,T) = W_0(\rho) + \gamma(\rho) U(\rho,T)$. The only ingredient missing to generate a full equation of state expressed as $U = U(\rho,T)$ and $W = W(\rho,T)$ is a relation between temperature and potential energy or virial along an isochore.

Rosenfeld and Tarazona \cite{ros98} derived from density functional theory an expression for the potential energy on an isochore:
\begin{eqnarray}
   U(\rho, T) = U_0(\rho) + \alpha(\rho) T^{3/5}. \label{Eq:URosenfeld}
\end{eqnarray}
Rosenfeld and Tarazona noted that the expression`` ... provides a good estimate of the thermal energy (...) near freezing densities only for predominant repulsive interactions'', and confirmed their expression by simulations of IPL potentials with different exponents. Equation~(\ref{Eq:URosenfeld}) has since been shown, however, to be an excellent approximation also for several models with attraction, including the KABLJ mixture.\cite{sci00,urp10,col00} This was regarded as a bit of a mystery, but can now be understood as a consequence of the fact that dynamics and heat capacity of strongly correlating liquids are well reproduced by a corresponding IPL system\cite{urp10} -- a consequence of the hidden scale invariance discussed briefly in the introduction (see Papers I-III).

Figure~\ref{Fig:URosenfeld} tests the Rosenfeld-Tarazona prediction  for the systems investigated in Fig.~\ref{Fig:UWischores}. The  prediction works very well, but less so for the supercooled Wahnstr\"om BLJ system, for which the formation of extended ``Frank-Kasper'' clusters at low temperatures affects the temperature dependence of the potential energy. \cite{UrpWahnstrom}

Combining Eq.~(\ref{Eq:URosenfeld}) with our results on isomorphs and isochores we can construct an equation of state for generalized LJ systems. The idea is to map any state point $(\rho,T,U,W)$ to the corresponding isomorphic state point  $(\rho_*,T_*,U_*,W_*)$ on a reference isochore (where it is implecitely understood that the parameters  ($W_0, \gamma, U_0, \alpha$) are evaluated at the reference isochore $\rho_*$):
\begin{eqnarray}
   T_*(\rho,T) &=& T \tilde\rho^{-\gamma}, ~~~ \tilde\rho \equiv \rho/\rho_*, \\
   U_*(\rho,T) &=& U_0 + \alpha (T \tilde\rho^{-\gamma})^{3/5},  \label{Eq:Ustar} \\
   W_*(\rho,T) &=& W_0 + \gamma U_*(\rho,T). \label{Eq:Wisochor}
\end{eqnarray}
From ($U_*,W_*$) we can calculate the contribution to the potential energy from the two IPL terms of the potential (compare Eqs.~(\ref{eq:Um}) and(\ref{eq:Un})):
\begin{eqnarray}
   U_{m}^*(\rho,T) 
                   &=& \frac{~3W_0 - (n - 3\gamma)\left( U_0 + \alpha (T \tilde\rho^{-\gamma})^{3/5}\right)}{m-n},~~~\label{Eq:UmsEOS}\\
   U_{n}^*(\rho,T)
                   &=& \frac{-3W_0 + (m - 3\gamma)\left( U_0 + \alpha (T \tilde\rho^{-\gamma})^{3/5}\right) }{m-n}.~~~
\end{eqnarray}
Finally, we use isomorphic scaling (Eqs.~(\ref{Eq:IsomorphU}) and (\ref{Eq:IsomorphW})) to go back to the $(\rho,T,U,W)$ state point:
\begin{eqnarray}
   U(\rho,T) &=&\tilde\rho^{m/3} U_{m}^* (\rho,T) + \tilde\rho^{n/3}U_{n}^*(\rho,T), \label{Eq:isomorph_scaling_U} \\ 
    W(\rho,T) &=&\frac{m}{3}\tilde\rho^{m/3} U_{m}^*(\rho,T)  + \frac{n}{3}\tilde\rho^{n/3}U_{n}^*(\rho,T). \label{Eq:isomorph_scaling_W}
\end{eqnarray}
The new equation of state given by Eqs.~(\ref{Eq:UmsEOS}) to (\ref{Eq:isomorph_scaling_W}) contains four parameters ($W_0, \gamma, U_0, \alpha$) evaluated at a reference density $\rho_*$. 
The parameters can be calculated from $U,W,C_{\rm V}^{\rm ex}$, and $\beta_{\rm V}^{\rm ex}$ at a single state point. A more convenient way of estimating the parameters might be: 
For a collection of $(\rho,T,U,W)$ state points, at least two of which are non-isomorphic, use isomorphic scaling (Eqs.~(\ref{Eq:IsomorphU})and~(\ref{Eq:IsomorphW})) to get the corresponding $(U_*,W_*)$ at the chosen reference density, and fit them to Eq.~(\ref{Eq:Wisochor}) to determine  $W_0$ and $\gamma$. Next, using the fitted value for $\gamma$, fit to Eq.~(\ref{Eq:Ustar}) in order to find $U_0$ and $\alpha$. 

To put the new equation of state to a test, the procedure described above was used to determine the four parameters ($W_0, \gamma, U_0, \alpha$) for the 12-6 KABLJ system using merely two state points. The resulting equation of state is shown as full lines in Fig. \ref{Fig:UWstate}, where it is compared to simulation results shown as data points. The agreement is good. The largest deviations are seen when changing density away from the reference density, most evident for the isotherm in the $WU$ diagram and the isomorph in the $pT$ diagram. Better agreement is expected if the density dependence of $\gamma$ is taken into account, but such a correction will not be attempted here. To summarize, combining the isomorph theory with two further approximations, a state-point independent exponent $\gamma$ and the Rosenfeld-Tarazona expression, a realistic equation of state with four parameters is arrived at.

\section{Conclusions}

Paper IV demonstrated that strongly correlating liquids have isomorphs, i.e., curves in the phase diagram along which structure and dynamics to a good approximation are invariant in reduced units. The main new results in this paper are predictions specific to generalized Lennard-Jones systems and the supporting numerical evidence.

Starting from the invariance of structure along isomorphs, a prediction for the shape of these in the $WU$ phase diagram was presented and shown to agree well with simulation results. It was shown that for a given system, the isomorphs all have the same shape in the $WU$ phase diagram; they are simply scaled versions of each other. Furthermore, the isomorph shape depends only on the exponents $m$ and $n$, i.e., all systems with, say, 12-6 LJ interactions have same isomorphs in the $WU$ phase diagram. What differs from system to system is the values of the density and temperature along an isomorph.

Throughout Papers I-V the $WU$ phase diagram was referred to extensively. This particular phase diagram has not been used much before, but it is evident that for strongly correlating liquids several features stand out when presented in the $WU$ phase diagram. Prominent examples are that isochores are straight lines (with slope $\gamma$) and the existence of the ``master'' isomorph discussed above for generalized LJ systems. 

The class of strongly correlating liquids includes, we believe, all van der Waals liquids and metallic liquids (but excludes hydrogen-bonding, covalent, as well as strongly ionic liquids). More simulations, as well as experiments, are needed to substantiate that the class of strongly correlating liquids is this large, but preliminary simulations of simple molecular models like\cite{sch08b} the asymmetric dumbbell and the Lewis-Wahnstrom OTP three-site model are encouraging. In particular, the isomorph properties of more complex molecule models need to be simulated, as well as theoretically contemplated; the scaling properties of molecules with fixed bond lengths are not trivial. It should be noted that even a quite complex system like a phospholipid membrane can be strongly correlating with respect to its slow degrees of freedom,\cite{ped10a} the interactions of which are dominated by van der Waals forces. Another interesting question is to study in detail the properties of liquids where the exponent $\gamma$ varies significantly throughout the phase diagram, for instance the  Weeks-Chandler-Andersen cutoff variant of the generalized binary LJ liquid that was recently investigated om detail by Coslovich and Roland, \cite{coswca} as well as Berthier and Tarjus.\cite{Berthier} Another system that deserves further investigation is the 200-100 LJ system\cite{200100} which we find to be an exception to the general observation that generalized LJ systems are strongly correlating -- presumably as a consequence of its extremely narrow range of interactions.

This series of papers investigated in depth the properties of strongly correlating liquids. We believe to have demonstrated that strongly correlating liquids are simpler than liquids in general. Hopefully this insight will turn out to be useful in furthering the understanding of the properties of liquids in general.

ACKNOWLEDGMENTS\\

URP  is supported by the Danish Council for Independent Research in Natural Sciences.
The center for viscous liquid dynamics ``Glass and Time'' is sponsored by the Danish National Research Foundation (DNRF).

\end{document}